\documentclass{aa}
\usepackage{amsmath}
\usepackage{natbib}
\usepackage{url}
\usepackage[english]{babel}

\usepackage{graphicx}
\usepackage{txfonts}
\usepackage{subfigure}
\begin{document}
   \title{Multiwavelength periodicity study of Markarian 501}

   \author{C. R\"odig
          \and
          T. Burkart
          \and
          O. Elbracht
          \and
          F. Spanier
        }

   \offprints{F. Spanier,  \email{fspanier@astro.uni-wuerzburg.de}}

   \institute{Lehrstuhl f\"ur Astronomie, Universit\"at  W\"urzburg,
             Am Hubland, D-97074 W\"urzburg
             }

   \date{Received ...  ,accepted ...}


  \abstract
   {Active Galactic Nuclei are highly variable emitters of
    electromagnetic waves from the radio to the gamma-ray regime. This variability
    may be periodic, which in turn could be the signature of a binary
    black hole. Systems of black holes are strong emitters of
    gravitational waves whose amplitude depends on the binary orbital
    parameters as the component mass, the orbital semi-major-axis and
    eccentricity.
   }
   {It is our aim to prove the existence of periodicity of the AGN
    Markarian 501 from several observations in different wavelengths.
    A simultaneous periodicity in different wavelengths provides
    evidence for bound binary black holes in the core of AGN.  }
  {Existing data sets from observations by Whipple, SWIFT, RXTE
    and MAGIC have been analysed with the Lomb-Scargle method, the
    epoch folding technique and the SigSpec software.
  }
   {Our analysis shows a 72-day period, which could not be seen in
     previous works due to the limited length of observations. This
     does not contradict a 23-day period which can be derived as a
     higher harmonic from the 72-day period.  }
   {}


   \maketitle
%

\section{Introduction}

Active Galactic Nuclei (\emph{AGN}) have been a class of most
interesting astronomical sources for a long time, mainly for their
high variability on time scales from years down to minutes. Nowadays
these variations may be observed from radio up to gamma-rays ($>10$
TeV). In most cases the variation of the AGN do not show signs of
periodicity but a unpredictable switching between quiescent and active
states and flaring behaviour. The explanation of such behaviour
requires several different approaches in order to explain the very
different timescales. Most of the models make use of the source size
(the black hole radius) for the minimum time scales or the accretion
efficiency and outer scales for the maximum time scales.


Some sources show a periodicity on a timescale of several tens of days in the optical, X-ray or TeV data like Mrk 421, Mrk 501, 3C66A and PKS
2155-304 \citep{hayashida1998,1999ApJ...521..561L,2001ICRC....7.2630K,2001ICRC....7.2695O}, whereas
the long-term optical light curves from classical sources such as BL Lacertae, ON 231, 3C273,
OJ 287, PKS 0735+178, 3C345 and AO 0235+16 usually suggest (observed) timescales of
several years \citep[e.g.][]{1988ApJ...325..628S,1997A&AS..125..525F,2001A&A...377..396R}. Models invoked to describe aperiodic variability do not give rise to explanations for periodic variations. It is therefore necessary to develop new models. One of these models
is the binary black hole (\emph{BBH}) model, which is explained in detail in the
next section. It is worth noting that the origin of these quasi periodic objects (\emph{QPOs}) may be related to instabilities of (optically thin) accretion disks (cf. e.g. \cite{2008APh....28..508F} and citations within). We do not want to stress this point but concentrate on the BBH model to explain the variability.

The existence of BBHs seems plausible since the host
galaxies of most AGN are elliptical galaxies which originate from
galaxy mergers. The study of BBHs and their interactions
become of primary importance from both the astrophysical and
theoretical points of view now that the first gravitational wave detectors
have started operating; gravitational waves, while certainly extremely
weak, are also extremely pervasive and thus a novel and very promising
way of exploring the universe. For a review on BBHs the reader is referred to \citet{2003AIPC..686..161K}.

BBHs have already been detected \citep{bbh_chandra,bbh_vlbi, 1998tx19.confE.435S, 2003ChJAA...3..513R},
but most sources are far too distant to be separated in radio
or x-rays (for a detailed discussion of the emission mechanisms see \citep{2008ApJS..174..455B}. Using a multi-messenger technique (combined electromagnetic
and gravitational wave astronomy) it should be possible to identify
periodic AGN as BBH systems by means of their
gravitational wave signal on the one hand while on the other hand
x-ray and gamma-astronomy could easily identify possible targets for
gravitational wave telescopes.

The aim of this paper is to present a complete data analysis of
Markarian 501 in different wavelengths searching for periodicity. This
will allow for restricting possible black hole masses and the distance
between the two black holes. Taking into account the model of
\citet{rieger00}, where the smaller companion is emitting a jet with
Lorentz factor $\gamma$ towards the observer, we expect periodic
behaviour over the complete AGN spectrum.

This paper is organised as follows. In Section \ref{sec:general}  we review
a model used to explain the variability and present the equations used to determine the parameters of the binary system under study.
Then in Section \ref{sec:data} the data we used
and the experiments are described.
In Section \ref{sec:method} we describe methodology for data analysis
and in \ref{sec:results} we present the results of the analysis, discuss
our findings and draw some conclusions.

\section{\label{sec:general} The binary black hole model}

As discussed above, the presence of close supermassive binary black hole (SMBBH) systems has been
repeatedly invoked as plausible source for a number of observational
findings in blazar-type AGN, ranging from misalignment and precession
of jets to helical trajectories and quasi-periodic variability \citep[cf. e.g. ][]{2006MmSAI..77..733K}.

What makes the BBH model an unique interpretation is that it facilitates facts otherwise not possible:
(1.) It is based on quite general arguments for bottom-up structure formation, (2.) it incorporates helical jet trajectories observed in
many sources, (3.) it provides a reasonable explanation
for long term periodic variability, (4.) it can explain, to some extend, quasi-periodic variability
on different time-scales in different energy bands \citep[e.g. ][]{2007Ap&SS.309..271R}.

The nearby TeV blazar Mrk 501 $(z = 0.033)$ attracted attention in 1997, when the source
underwent a phase of high activity becoming the brightest source in the sky at TeV energies.
Recent contributions like \citet{rieger00, rieger01} have proposed a possible
periodicity of $P_\textrm{obs} = 23$ days, during the
1997 high state of Mrk 501,  which could be related to the orbital
motion in a SMBBH system.  The proposed model
relies on the following assumptions: (1.) The observed periodicity is associated with a relativistically moving feature
in the jet (e.g. knot).  (2.) Due to the orbital motion of the
jet-emitting (secondary) BH, the trajectory of the outward moving knot is expected to be a
long drawn helix. The orbital motion is thus regarded as the
underlying driving power for periodicity.  (3.) The jet, which
dominates the observed emission, is formed by the less massive BH.
(4.) The observed periodicity arises due to a Doppler-shifted
flux modulation which is caused by a slight change of the inclination angle with respect to the line of sight.

Following Rieger \& Mannheim (2000)(hereinafter RM), we assume
that the observed signal periodicity has a geometrical origin, being a consequence of a Doppler-shifted modulation.
It is therefore possible to relate the observed signal period
$P_{obs}$ to the Keplerian orbital period
\begin{eqnarray}
\label{keplerperiod}
\Omega_{k} = \frac{\sqrt{G (m + M)} }{d^{3/2}},
\end{eqnarray}
by the equation \citep[cf.][]{1992A&A...255...59C,1994A&A...290..357R}
\begin{eqnarray}
\label{periodrelation}
P_{obs} = (1+z) \left( 1 - \frac{v_{z}}{c} \cos i \right) P_{K} ,
\end{eqnarray}
where  $v_{z}$ is the typical jet velocity assumed to be $v_{z} = c \left( 1-\gamma_{b}^{-2} \right)^{1/2}$.
Current emission models \citep[e.g.][]{1999APh....11...59S} favour an inclination angle $ i \approx 1/\gamma_{b}$ ,
with typical bulk Lorentz factors in the range $10-15$ \citep[e.g.][]{1996A&A...315...77M,1999ApJ...511..136S}.

For a resolved emission region (e.g. a blob
of plasma) with spectral index $\alpha$, the spectral flux modulation by Doppler
boosting can be written as
\begin{eqnarray}
\label{dopplerboosting}
S(\nu) = \delta(t)^{3+\alpha} S^{\prime} (\nu),
\end{eqnarray}
where $S^\prime$ is the spectral flux density measured in the co-moving frame and  $\delta(t)$
denotes the Doppler factor given by
\begin{eqnarray}
\label{dopplerfactor}
\delta(t) = \frac{1}{\gamma_{b} \left[ 1 - \beta_{b} \cos \theta(t)\right]},
\end{eqnarray}
where $\theta(t)$ is the angle between the direction of the emission region (e.g. blob) and the line of sight.
Due to the orbital motion around the centre-of-mass,
the Doppler factor for the emission region is a periodical function of time. In the simplest case (without determination) the Doppler
factor may be written as follows (RM)
\begin{eqnarray}
\label{dopplerfactor}
\delta(t) = \frac{\sqrt{1-\left(v_{z}^{2} + \Omega^{2}_{k} R^2 \right)/c^2}}{1 - \left(v_{z} \cos i + \Omega_{k} R  \sin i \sin \Omega_{k}t\right)/c } .
\end{eqnarray}
A periodically changing viewing angle may thus naturally lead to
a periodicity in the observed lightcurves even for an intrinsically constant flux.

Depending on the position of the jet-emitting BH along its orbit, the
Doppler factor has two extremal
values, so that one obtains the condition
\begin{eqnarray}
\label{maxmincondition}
\delta_\textrm{max}/\delta_\textrm{min} \approx  f^{1/(3+\alpha)},
\end{eqnarray}
where $f = S_\textrm{max}(\nu)/S_\textrm{min}(\nu)$ is the observed maximum to minimum ratio.
 Consequently, by inserting Eq. (\ref{dopplerfactor}) in Eq. (\ref{maxmincondition})
 one finds
\begin{eqnarray}
\label{com-distance}
\Omega_{k} R = \frac{f^{1/(3+\alpha)}-1}{f^{1/(3+\alpha)}+1} \left( \frac{1}{\sin i} - \frac{v_{z}}{c} \cot i \right) c.
\end{eqnarray}
Finally,  by using the definition of the Keplerian orbital period Eq. (\ref{keplerperiod}) and noting that $R = M d / (m + M)$ one derives the binary mass ratio equation (RM)
\begin{eqnarray}
\label{mass-ratio}
\frac{M}{(m + M)^{2/3}} &=& \frac{P_{obs}^{1/3} \left(\sin i \, \right)^{-1} c}{\left[\left(1+z\right) 2 \pi \, G \right]^{1/3}} \frac{f^{1/(3+\alpha)}-1}{f^{1/(3+\alpha)}+1} \left(1- \frac{v_{z}}{c} \cos i \right)^{2/3},
\end{eqnarray}
and from Eq. (\ref{periodrelation}) we get
\begin{eqnarray}
\label{mass-ratio2}
m + M &=& \left( \frac{2  \pi (1+z)  \left( 1 - \frac{v_{z}}{c} \cos i \right)}{P_{obs}} \right)^2 \frac{d^3}{G}.
\end{eqnarray}
To be able to determine the primary and secondary black hole masses $m$ and $M$ by solving simultaneously
Eq. (\ref{mass-ratio}) and (\ref{mass-ratio2}) we have to provide an additional assumption or a third equation. RM obtained the binary separation $d$ by equating the time-scale
for gravitation radiation with the gas dynamical time-scale \citep{1980Natur.287..307B}. We have not adopted this
condition since it is somewhat arbitrary to assume $T_{gas} = T_{GW}$, nowadays. Therefore we will leave the distance
$d$ as a free parameter in the equations. Instead, we solve Eq. (\ref{mass-ratio}) and (\ref{mass-ratio2}) by adopting reasonable values for the total binary mass from observational findings, $M+m=10^{8.62}$ solar masses \citep{2005ApJ...631..762W}.

As outlined, we can determine the orbital parameters of the system in case of a perfectly circular orbit. Notably,
the assumption of a perfectly circular orbit is simplifying the problem to some extend. It is true that the orbits
tend to circularise due to gravitational radiation, but this happens within a time-scale of the same order of magnitude
as the merging time-scale \citep{1963PhRv..131..435P,1987MNRAS.224..567F}. Therefore it is possible that the
constituting black holes of a binary system are still on eccentric orbits, in which case the method by RM will be only
an approximation to the real situation. Recently \cite{2002A&A...388..470D,2003A&A...410..741D} extended the model of RM to elliptical orbits
including the eccentricity $e$, which reduces to the circular model in case of $e \to 0$.
In case of an eccentric orbit a degeneracy in the solutions appears making it impossible to determine the orbital
parameters in an unique fashion. As pointed out further by \cite{2002A&A...388..470D}  the degeneracy can only be solved by
the detection of a gravitational wave signal.

\section{\label{sec:data}Data}

Data was taken from a variety of different sources, earthbound telescopes like MAGIC or WHIPPLE, but also from satellite experiments like RXTE and SWIFT. 
In contrast to any previous studies on this source, we explicitly excluded the time of the flare in 1997, so to determine whether a periodicity would still be 
present in the quiescent state of Mrk 501.
We didn't make use of any raw data material, but only used data that was background-corrected and processed by the respective collaborations
, to which we refer for more details. Therefore the units 
of the plots are arbitrary, a fact that doesn't cause any problems for our analysis as we are not interested in any absolute flux values of the source.

\subsection{RXTE}
The Rossi X-ray Timing Explorer (RXTE) is an orbital X-ray telescope for the observation of black holes, neutron stars, pulsars and X-ray binaries. 
The RXTE has three instruments on board: the Proportional Counter Array (PCA, 2 - 60 keV), the High-Energy X-ray Timing Experiment (HEXTE, 15 - 250 keV) and the 
so-called All Sky Monitor (ASM, 2 - 10 keV). The PCA is an array of five proportional counters with a time resolution of $1 \mu s$ and an angular of $1^{\circ}$  FWHM, 
the energy resolution amounts to 18\% at 6 keV. For HEXTE the time sampling is $8 \mu s$ with a field of view of $1^{\circ}$  FWHM at an energy resolution of 15\% at 
60 keV and the ASM consists of three wide-angle cameras which scan 80\% of the sky every 90 minutes \citep{1996SPIE.2808...59J, rxtewebsite}. 

\begin{figure}
 \includegraphics[width=0.95\linewidth]{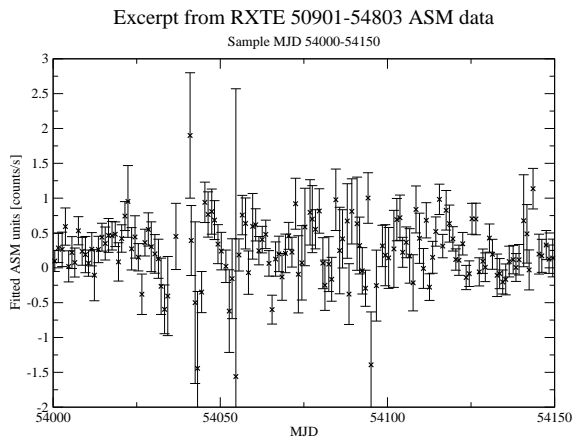}
 \caption{\label{rxtedata} Subset of the whole available ASM data on Mrk 501. The sampling is not equidistant and the data points clearly differ significantly from zero.}
\end{figure}

The data from RXTE we used was gathered by the ASM from March 1998 until December 2008 and a sample from the whole data set can be seen in Fig.~\ref{rxtedata}. 
We are only plotting a subset to show the actual sampling structure of the RXTE data, firstly because the data is extremely plentiful 
(it contains 3697 data points) and secondly to have a direct comparison to the SWIFT data in Fig.~\ref{ausschnittswift}. The data shows a rms of
$5.15 \cdot 10^{-1}$ (in arbitrary units) with an average signal-to-noise ratio of $1.22$. 

\subsection{SWIFT}
The SWIFT satellite is part of the SWIFT Gamma-Ray Burst Mission planned and executed by NASA dedicated to the study of GRBs and their afterglows. 
It consists of three instruments, the Burst Alert Telescope (BAT) which detects gamma-rays in a range from 15 to 150 keV with a field of view of about 2 $sr$ and 
computes their position on the sky with arc-minute positional accuracy. 
After the detection of a GRB by the BAT, its image and spectrum can be observed with the X-Ray Telescope (XRT, 300 eV - 10 keV) and the Ultraviolet/Optical Telescope 
(UVOT, 170 - 650 nm) with an arc-second positional accuracy \citep{2003AAS...202.4804B, swiftwebsite}.

\begin{figure}
 \includegraphics[width=0.95\linewidth]{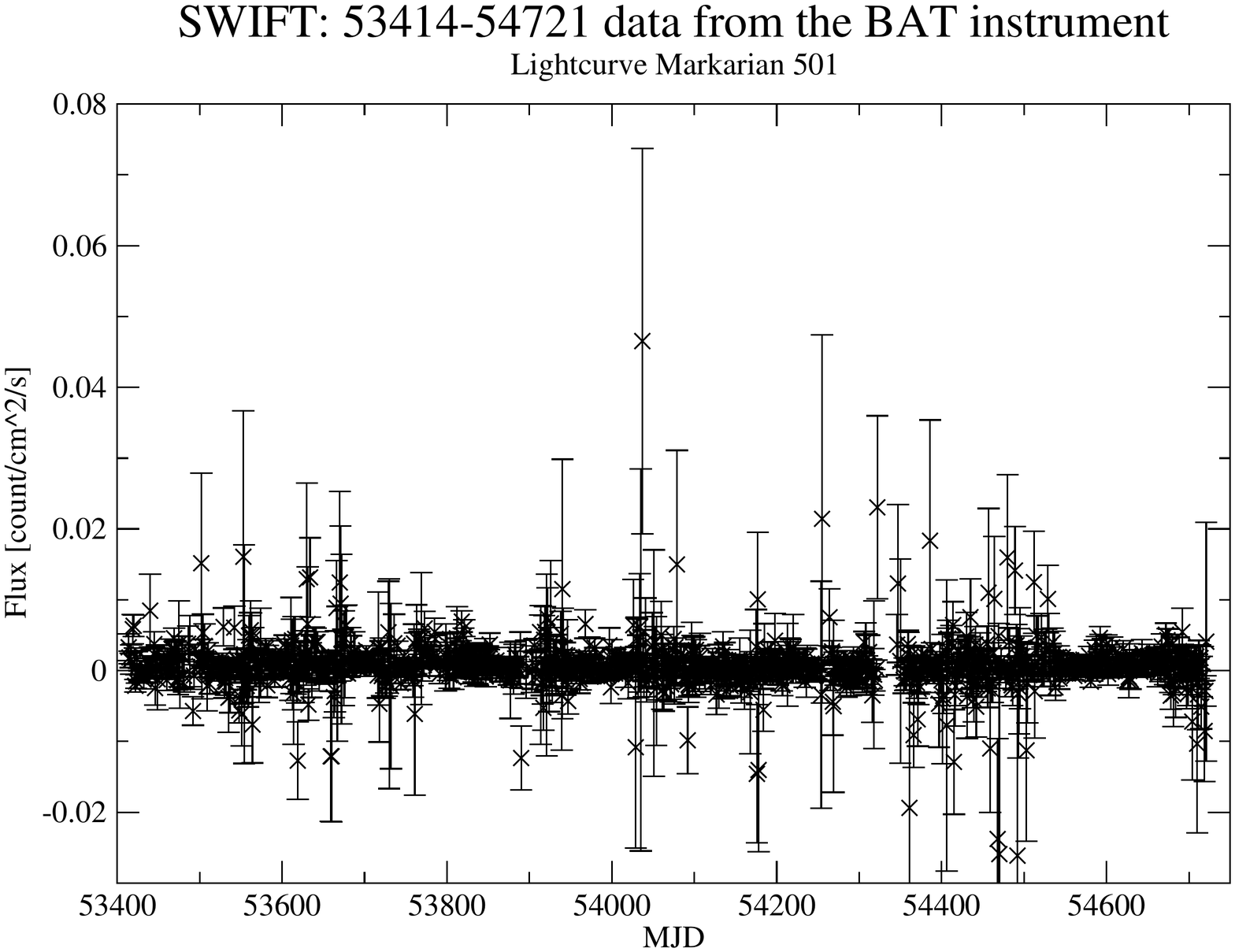}
\caption{\label{swiftdatapure}Complete available data for SWIFT containing 1153 data points with error bars. Background has already been subtracted before data retrieval (for details see \citep{swiftremove}).} 
\end{figure}

\begin{figure}
 \includegraphics[width=0.95\linewidth]{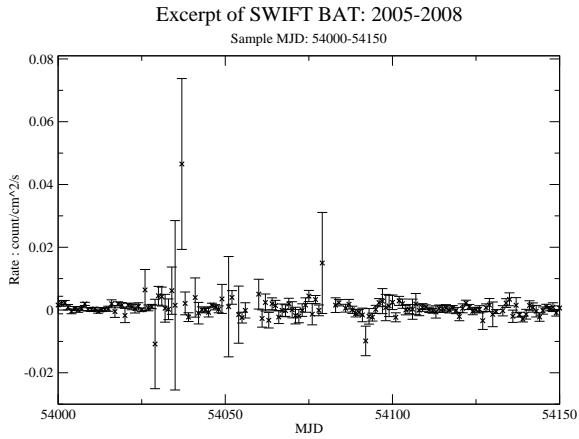}
\caption{\label{ausschnittswift}Subset of SWIFT data around the maximum amplitude in order to show the single points more clearly. The highest peak lies 12.8 $\sigma$ above the mean flux and is therefore considered a fluctuation. So the remaining of the lightcurve is believed to consist merely of zeroes.}
\end{figure}

The data that was used can be obtained from a database at the website of the NASA \citep{swifturl}. We made use of data from February 2005 to September 2008. 
As can be seen in Fig.~\ref{swiftdatapure}, the lightcurve provided by the SWIFT satellite looks very much like background noise. 
Especially when comparing the size of the error-bars to the amplitude of the signal in the subset in Fig.~\ref{ausschnittswift}, 
common sense implies that this lightcurve should be regarded with caution. The subset was chosen around the maximum amplitude measured 
within the whole sample and the same time-span was then plotted for RXTE in Fig.~\ref{rxtedata} for comparison. The rms of the flux 
of the whole data sample is $35.7 \cdot 10^{-4}$ (in arbitrary units) with an average signal-to-noise ratio of only $0.570$. 
As a result of this, we consider the data from SWIFT merely as an indicator, but not as solid evidence that seriously influences the outcome of the hypothesis tests.


\subsection{MAGIC}
\begin{figure}
 \includegraphics[width=0.95\linewidth]{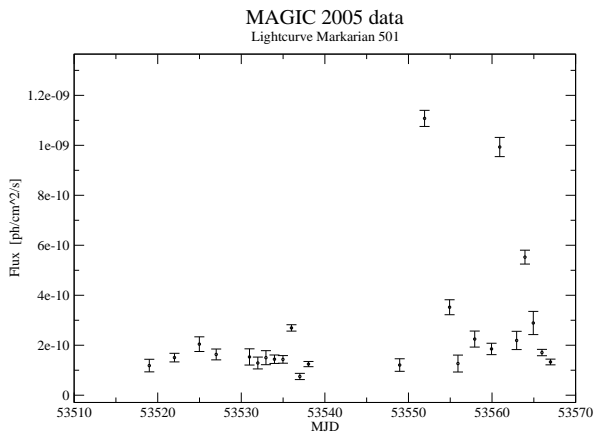}
\caption{\label{magicdata}Measured High-Energy data from MAGIC in 2005, the set contains 24 data points.}
\end{figure}

MAGIC is an Imaging Air Cherenkov Telescope (IACT) situated at the Roque de los Muchachos on one of the Canary Islands, La Palma. It 
has an active mirror surface of 236 m$^2$ and a hexagonal camera with a diameter of 1.05 m consisting of 576 photomultipliers. The 
trigger and analysis threshold are 50 - 70 GeV and the upper detection limit is 30 TeV, for further details on the MAGIC telescope 
see \citet{magic1999}. The data we used was collected from May to July 2005 in the energy 
range of 150 GeV to 10 TeV \citep[]{magicdatapaper} and can be seen in Fig.~ \ref{magicdata}. It contains of 24 data points with a
rms of $2.63 \cdot 10^{-10}$ (in arbitrary units) and an average signal-to-noise ratio of $11.0$.

%

\subsection{WHIPPLE}
\begin{figure}
 \includegraphics[width=0.95\linewidth]{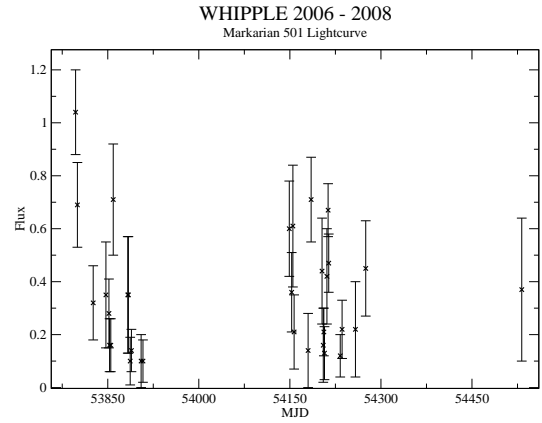}
\caption{\label{whippledata}Preliminary High-Energy data measured by WHIPPLE from 2006 to 2008, the set contains 33 data points}
\end{figure}

In the early 1980s, the 10 m Whipple Telescope was built at the Fred Lawrence Whipple Observatory in southern Arizona. It is 
still fully operational and is nowadays used for long-term monitoring of some interesting astrophysical sources in the energy 
range from 100 Gev to 10 TeV. The data of Mrk501 we used was taken from March 2006 to March 2008 and is preliminary \citep{whipple}.  It shows a rms of 
$2.33 \cdot 10^{-1}$ (in arbitrary units) and an average signal-to-noise ratio of $2.45$ (see Fig. \ref{whippledata}).

\section{\label{sec:method}Analysis technique}
In observational sciences the observer can often not completely
control the time of the observations, but must simply accept a certain
dictated set of observation times $t_i$.  In order to process unevenly
sampled data, simple Fourier-transformations cannot be used, but must be
replaced by a more elaborate analysis. There are several ways of how to tackle this problem. In our work we made use of
three different methods. These we selected by the criterion of being mutually fairly orthogonal so as to reduce systematic errors. In the following paragraphs two of them
will be outlined briefly, the third was gratefully accepted from the Dept.  for
Astroseismology of the University of Vienna.

\subsection{Lomb-Scargle and Epoch Folding}
\subsubsection{Lomb-Scargle}
A full derivation of the Lomb normalised periodogram (spectral power
as a function of angular frequency $\omega \equiv 2\pi f > 0) $ may be
found in \citet{lomb} and \citet{scargle} . The algorithm transforms
a discretely and unevenly sampled set of data points into a
quasi-continuous function in Fourier space, essentially by applying a
$\chi ^2$-test on an equivalent to linear least square fitting to the
model
\begin{eqnarray}
 h(t)= A \cos \omega t + B \sin \omega t
\end{eqnarray}
In order to detect low frequency signals, the oversampling method
\citep{recipes} is usually considered. However, spurious effects are
thus introduced and particular attention is necessary in analysing
the resulting periodogram \citep[see also][]{vaughan}. An ill-chosen value of the oversampling parameter is found to produce
artificial oscillations in the periodogram that can easily
be interpreted wrongly. We used the oversampling method and in Sec.~\ref{sec:blubber} describe  means of
controlling these oscillations and we depict a way of statistically
evaluating them as one of the major contributions to the uncertainty
of an identified period.

\subsubsection{Epoch Folding}
Following \citet{leahy}, the epoch-folding statistic is defined as
\begin{eqnarray}
 \chi ^2 (P) = \sum_{j=1}^{M} \frac{ \left( \bar{x}_j - \bar{x}\right) ^2}{\sigma_j^2}
\end{eqnarray}
where $M$ denotes the number of phase bins and $\sigma_j^2$ the
population variance of $\bar{x}_j$. A high $ \chi^2$ value ($\gg M-1$)
will signal the presence of a periodicity for which the significance
can be estimated from the $\chi^2_{M-1}$ distribution \citep[cf.][]{larsson,schwarzenberg}. Therefore, this method
yields an estimate of whether or not the data is distributed
Gaussianly; in the case of coloured noise being part of the
experimental error, as is the case in the present study, the method
may not yield reliable results, as the noise itself, if the signal to
noise ratio is too bad, could cause the distribution of the data to be
non-Gaussian. 
We used this method especially in order to cross-check the results of the other algorithms for artifacts. This means, whenever a periodogram shows high values in Fourier power, we construct the phase diagram of the corresponding period and try fits onto  formulae of periodic signals, like: \begin{eqnarray} y = A_0+A_1\cdot \sin(A_2\cdot x+A_3) \label{epochfit}\end{eqnarray} where the $A_i$ are the degrees of freedom in the fit. We also tried higher derivatives or higher powers of trigonometric functions, however the formula Eq.~\ref{epochfit} proved to yield the best results.

\subsection{SigSpec}
This software \citep[for details see][]{ReegenSigspec} presents a combination of the techniques mentioned
above, and has many more other features in addition. One aspect that needs to be
mentioned is that where the Lomb-Scargle Periodogram is presented as a
continuous function of the period, SigSpec only provides discrete
values of those periods, that are above a certain threshold-level of spectral significance. The latter is the inverse False-Alarm-Probability $ P(>y)=1-(1-\exp(-y))^M \label{falsealarm}$
 scaled logarithmically (specified significance threshold $y$ and number of independent frequencies $M$). The conversion of a chosen threshold for maximum spectral significance into its corresponding ``individual'' spectral significance threshold for accepting detected peaks is very similar to \citet{scargle} and can be found in \citet{ReegenSigspec}.
SigSpec cannot deal with coloured noise, however it is possible to choose different spectral significance thresholds for different frequency regions \citep{ReegenSigspec}.
\subsection{\label{sec:blubber} Random-noise-modelling and Random-sampling}
In order to handle the uncertainties of the results we used two approaches as follows. One way of testing the significance of the resulting
Fourier $\leftrightarrow$ Lomb-Scargle-transformation is
``Random-sampling'': we randomly choose a certain number $P$ of data points
from the original data set thus creating $n$ subsets, transform all of
them (using one of the three algorithms at a time) individually and add the thus acquired $n$ Fourier-sets together so to form one single
periodogram. $P$ was chosen to be $2/3$ of the original length of the lightcurve so as to guarantee sufficient degrees of freedom for permutation  whereas at the same time not forfeiting the low frequency detection for the short data sets, which gets the more difficult the shorter the subset becomes.
 In order to facilitate proper summation of the results, the transformations of the $n$ subsets need to have the same frequency resolution and frequency window so one really sums up the function at equal points.
  If the resulting periodogram still shows appreciably high
$\chi ^2$ values, it can be deduced that the corresponding
periods are actually physically relevant. If otherwise the peaks
become increasingly broadened, it is rather likely that they do not
correspond to a physically relevant period but stem from noise and/or
aliasing effects.

\begin{figure}
\includegraphics[width=1.0\linewidth]{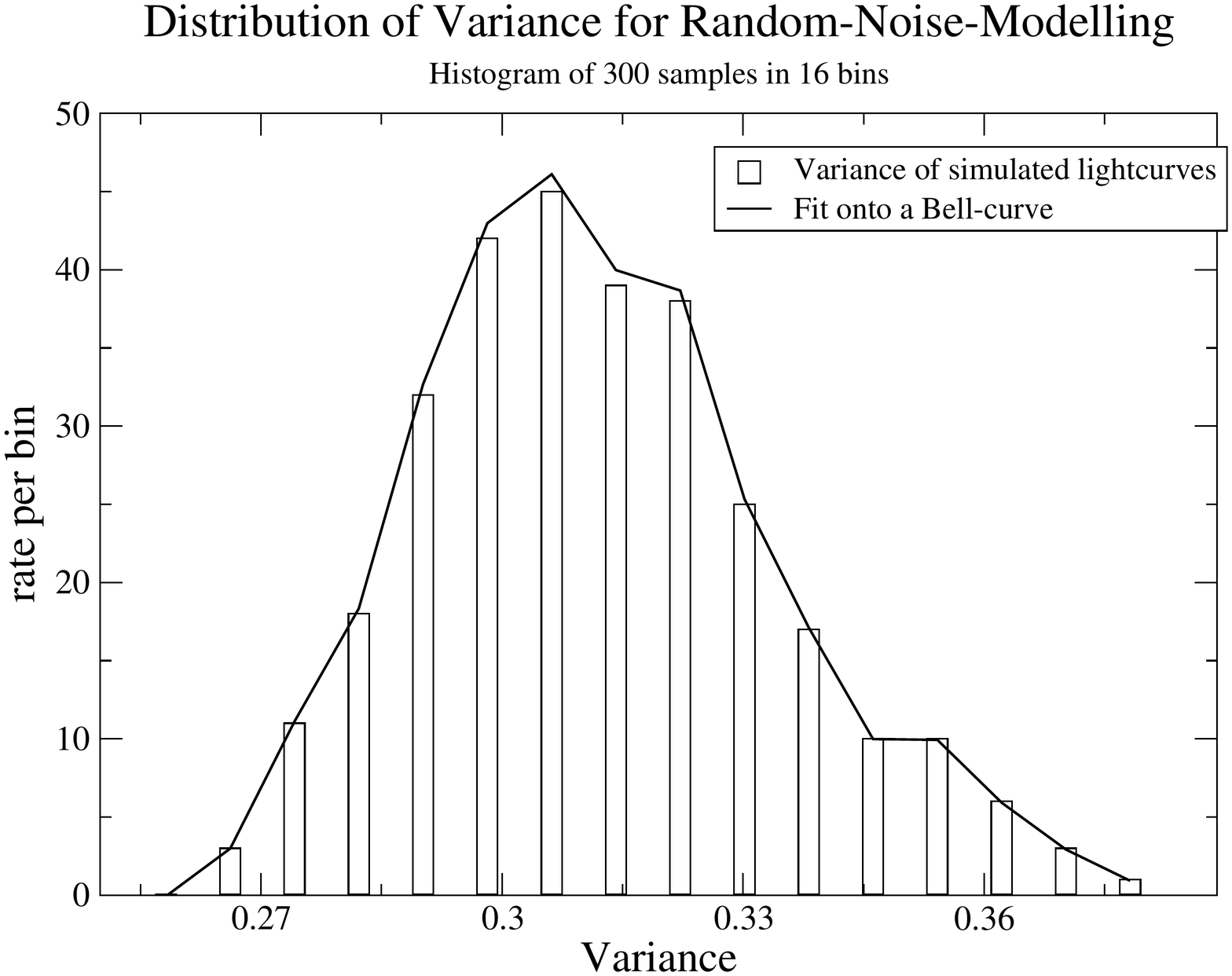}
\caption{\label{histotest.pdf} Histogram of the variance of $300$ Random-noise-modelling samples and the one-parameter fit onto a Gaussian-Distribution yielding a correlation coefficient of 0.975010}
 
\end{figure}

Another method is ``Random-noise-modelling'': We take the data points $x_i$ and their error-bars $\delta_i$, create a uniformly distributed random  number $\xi_i \in\, [-\delta_i,+\delta_i]$  and add $\xi_i$ to the corresponding $x_i$. The idea behind this is to weigh the influence of a bad signal-to-noise ratio. If a data set has good signal-to-noise ratio, the random-noise-modelling will scarcely effect the resulting periodogram at all as the interval $\pm \delta_i$  is small and therefore the new set, created by the Random-noise-modelling is almost identical to the original one. If, however, the $\delta_i$ are large, i.e. the data has a bad signal-to-noise ratio, the new set will differ significantly from the old and it is to be expected that the period-analysis will yield different results. So we have found a way of assigning tendentially larger uncertainties to the results from data sets of bad quality . It might be argued that the distribution of the $\xi_i$ should be Gaussian, other than uniform, however, we preferred the uniform distribution as it leads to larger differences of the new sets and the originals and gives therefore a larger uncertainty in the final results, which we are otherwise underestimating. The variance of the new set is statistically converging towards that of the original as for each data-point $x_i$ the $\delta_i$ and from that the $\xi_i$ is calculated.  The distribution of the variance is given in Fig.~\ref{histotest.pdf}. This graph shows that the distribution of the variance, and thus of the rms of the simulated lightcurves, is gaussianly distributed around the original variance. We fitted with the formula 
\begin{eqnarray}y = \frac{1}{N\,   \sqrt{2\,\pi}\, A_0}\times\exp\left(\frac{-(x-\mu)^2}{(2\, \pi\, A_0^2)}\right)\end{eqnarray}
where $\mu=0.306117$ is the mean of the original set and $N=0.030765$ a constant for normalisation. The fit yielded  $A_0=0.009955$ with a correlation coefficient of $0.975010$.   Thus we are left with a new lightcurve, that will, statistically, have similar rms and variance as the original one. We, again, pick $n$ of the new lightcurves and execute the transformation.



Throughout the analysis we took $n =\, 3$ representative samples, each.

The results are accumulated (= added up) onto one final periodogram, which provides the means of determining the
significance level $\sigma$ of the whole analysis. If now the
periodogram resulting from the summation should exhibit its highest
peak at places where the underlying single curves show nothing but a
forest of little noise peaks, we deduce that the corresponding period
is merely due to noise and has no physical meaning whatsoever. This
effect is mainly observed for very small periods ($P < 10$ days),
whereas for higher periods the displacement of the large peaks along
the $period$-axis dominate.

Both random-methods were applied to all data sets independently, i.e. the original sets were taken, for each of them we created $3$ sets using the Random-noise-modelling and (again from the original) $3$ subsets using the random-sampling. The analysis of
the data and its random models was carried out respecting all three
numerical methods (see above). The simulated light curves are treated on equal footing as the original set, no weighing is applied, nor are the curves mixed. Only the very final periodograms (Fig. ~\ref{rxtelomb}) present the sum of all $7$, individually analysed, lightcurves for each experiment. As shown in Fig. ~\ref{veritas} the Lomb-technique for high oversampling parameters may display oscillations, which appear to be Gaussian-shaped and more or less symmetrical around some central period. We believe these oscillations to be a mere artifact due to choosing the oversampling parameter slightly too big, a choice which is however necessary in order to resolve larger periods at all. Therefore we chose to fit enveloping Gaussian curves onto these structures.  The $\sigma$ levels were extracted from
the HWHM value of the envelope, that was fitted onto the final
(accumulated and normalised) periodograms.

For the SigSpec tool, however, this formula is not applicable and also the method of fitting the envelope is generally underestimating the uncertainty.
According to \citet{kataoka} the normalised power spectral density
exhibits a power-law $P(f) \propto f^{-1.0}$ for low (pink) frequencies, and
the transformations were consequently corrected thus ("pink corrected").  In order to
avoid spurious signatures from gaps, the data sets were each examined
for their sampling structure so as to treat peaks arising around the
positions of recurring data-gaps with additional caution.

\section{\label{sec:results}Results and Discussion}

\begin{figure}
\includegraphics[width=0.95\linewidth]{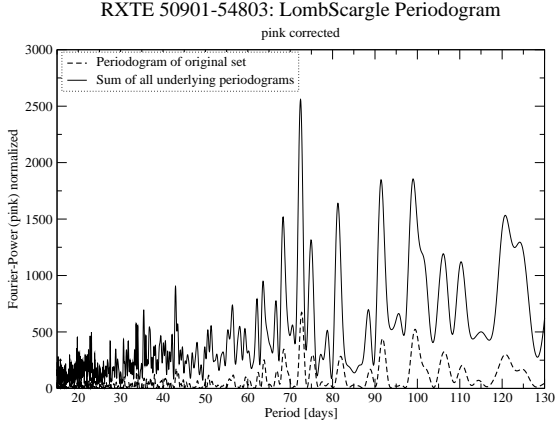}
  \caption{\label{rxtelomb} Pink-corrected Lomb-Scargle Periodogram for the RXTE data: 1998-2008  of Mrk 501. This final periodogram is the sum of the periodograms of the original lightcurve and the simulated lightcurves. The dashed graph is the periodogram of the original data set}
\end{figure}

\begin{figure}
 \includegraphics[width=0.95\linewidth]{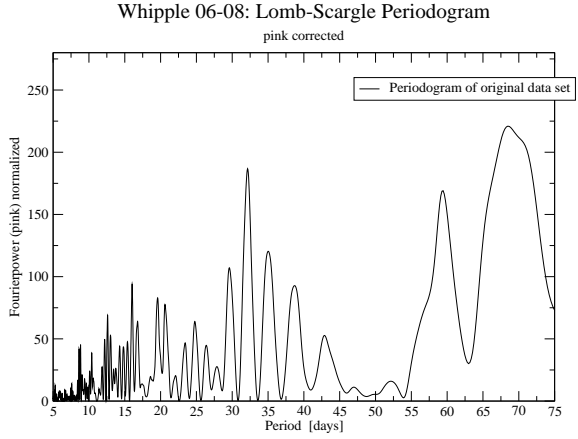}
  \caption{\label{veritas} Pink-corrected Lomb-Scargle Periodogram for the preliminary WHIPPLE data 2006-2008 of Mrk 501. This final periodogram was produced using a very high oversampling parameter so as to resolve long periods in this short data set. Thus the influence of artifacts dominates for periods larger than 30 days }
\end{figure}

\begin{figure}
 \includegraphics[width=0.95\linewidth]{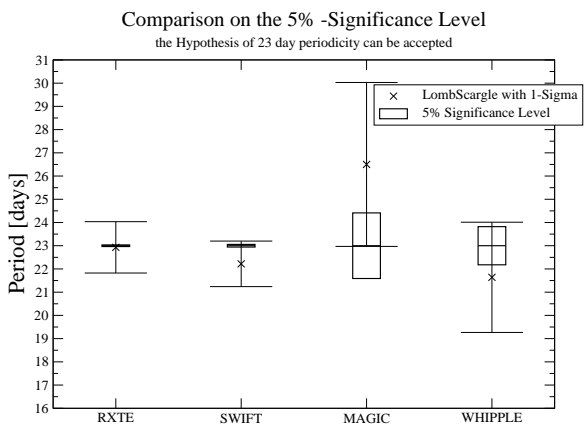}
  \caption{\label{comp23} The two sided-hypothesis test of whether the $23$-day period can be accepted within the $5\% $-level for the Lomb-Scargle algorithm: All found periodicities are consistent on the $5\%$ confidence level. }
\end{figure}

\begin{figure}
 \includegraphics[width=0.95\linewidth]{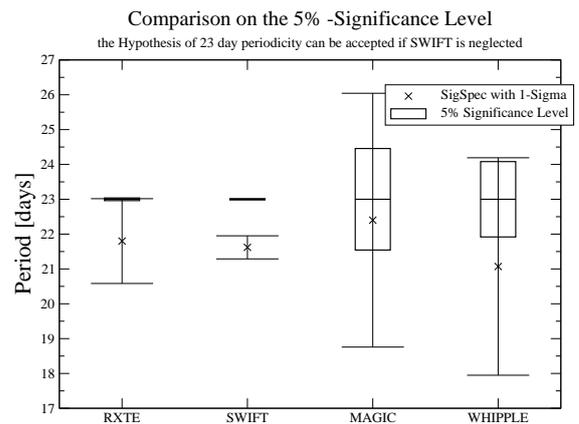}
  \caption{\label{compsig23} The two sided-hypothesis test of the $23$-day period for the SigSpec algorithm: The SWIFT data does not support the hypothesis, however, we very much doubt the quality of that data set and consider the hypothesis as valid on the $5\%$-level.}
\end{figure}

\begin{figure}
 \includegraphics[width=0.95\linewidth]{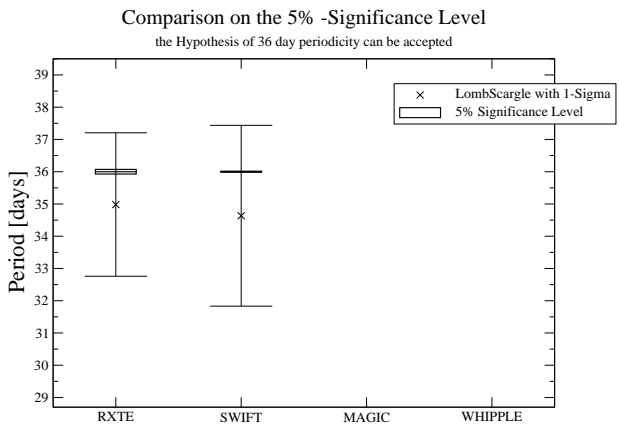}
  \caption{\label{comp36} The two sided-hypothesis test of the $36$-day period for the Lomb-Scargle algorithm: the frequencies of the first harmonic are consistent on the $5\%$-level. The data sets from the experiments, which are not included, were too short for the algorithm to produce reliable results. }
\end{figure}

\begin{figure}
 \includegraphics[width=0.95\linewidth]{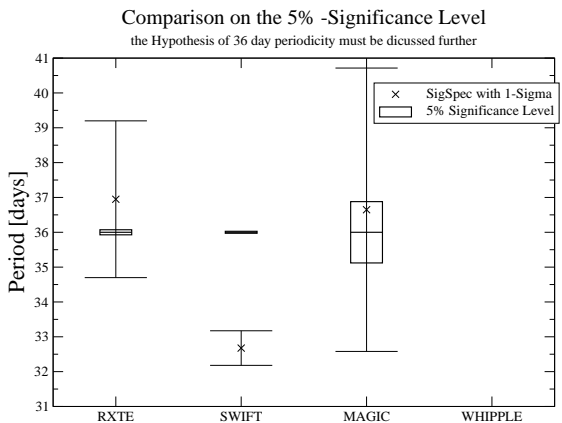}
  \caption{\label{compsig36} The two sided-hypothesis test of $36$-day period for the SigSpec algorithm: This  period can be found to be persistent in the data on a $5\%$-level, if again SWIFT is not taken into account}
\end{figure}

In this section we present the results of our analysis methods applied
to the various data sets available for Markarian 501. First of all we show
the periodograms calculated from the data (Figs.~\ref{rxtelomb},~\ref{veritas}),
then we present the results of the hypothesis tests (in Figs.~\ref{comp23}~-~\ref{compsig36})
and summarise all this in Table~\ref{tab:compfinal}.

\begin{center}

\begin{figure}
\includegraphics[width=1.0\linewidth]{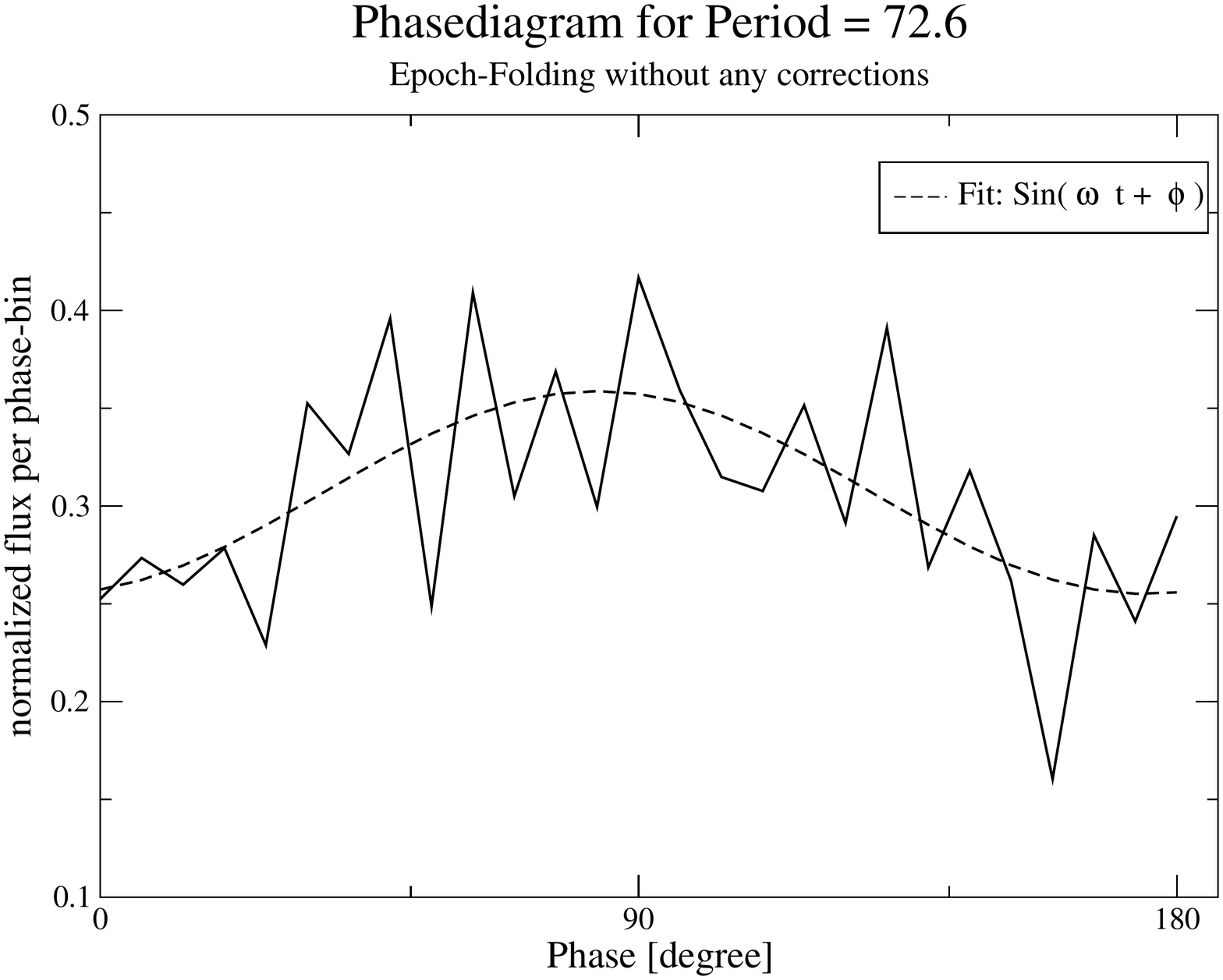}
\caption{\label{phasediagram} Phase Diagram of the period = 72.6 day in the RXTE 1998-2008 data: The Signature is clearly that of a sinusoidal signal.
\newline
The fit with the formula: $y = A_0+A_1\cdot \sin(A_2\cdot x+A_3)$ yielded $A_0 = 0.306917$, $A_1 = 0.051881$, $A_2 = 0.086545$, $A_3 = -0.466277$}
\end{figure}
\end{center}

In Fig.~\ref{rxtelomb} and Fig.~\ref{veritas} we show the typical
signatures of the fundamental frequency in RXTE and the first harmonic
in WHIPPLE, respectively. The periodogram of the preliminary WHIPPLE data was produced using a very high oversampling parameter, so as to resolve the long periods; however the data set is too short for the Lomb-Scargle algorithm to produce reliable results. In addition the Epoch-Folding-phase diagram in
Fig.~\ref{phasediagram} shows the signature of the $72.6$ day-period
to be clearly sinusoidal.  The overall hypothesis tests for both the
first and second harmonic of $72 \pm 4.3$ days are shown in Fig.~\ref{comp23}~-~\ref{compsig36}. The
hypothesis-tests are evaluated according to standard statistics using
two-sided tests on the specified confidence level.
The points with their error-bars are the identified periods with their 1-$\sigma$-uncertainties whereas the boxes show the area in which the hypothesis can be accepted, if overlapping. We find, that the 23-day period can be accepted on a $5\%-$level for the SigSpec results (cf. Fig.~\ref{compsig23}) and on a  $1\%-$level for the Lomb-Scargle results (cf. Fig.~\ref{comp23}) when all four experiments, that we consider significant, are compared. That means, that we accept a hypothesis even if the SWIFT results indicate the contrary (see Sec.~\ref{sec:data} for discussion).
For the Lomb-Scargle analysis of the 36-day period only the data-sets of sufficient length could be considered due to the fact that the detection of low frequencies requires longer time spans in the original data. Fig.~\ref{comp36} shows that a period of $1/2\cdot (72.2\pm 2.3)$ days can be considered present in the two data sets on a $5\%-$level.
In Fig.~\ref{compsig36}
we show that on a $5\%-$level, the first harmonic, $36$ days, as identified by the SigSpec algorithm is
still consistent within the 1-$\sigma$ uncertainties of the fundamental period, $1/2\cdot (72.6\pm 4.3)$ days,
even though we probably underestimated the uncertainties in our method for this algorithm. 
The phase diagram in Fig.~\ref{phasediagram} shows that  of the 4-parameter fit onto Eq.~\ref{epochfit} yields 0.621577 for the correlation coefficient, with an RMS per cent error of 0.18285 for the parameter $A_2 = 0.0865454$, which corresponds to a period of 72.6 days via the fundamental relation $\omega= 2 \pi /T$. We interpret this high correlation coefficient as a clear sign the periodicity itself to be of sinusoidal shape.

The approach to look for the harmonics of 72 days is motivated chiefly by Figs.~\ref{rxtelomb}
and \ref{phasediagram} for this period dominates over all other structures. However, the ASM
data is the only set, plentiful and long enough so the algorithms can detect and identify a
period clearly. Given the fact that all other samples are dominated either by noise or gaps
it is of great importance to seek for the persistent appearance of the harmonics throughout
all sets, even though a single detection of a period in one sole set would for itself scarcely be considered significant.

%

\begin{figure}[h]
\includegraphics[width=0.95\linewidth]{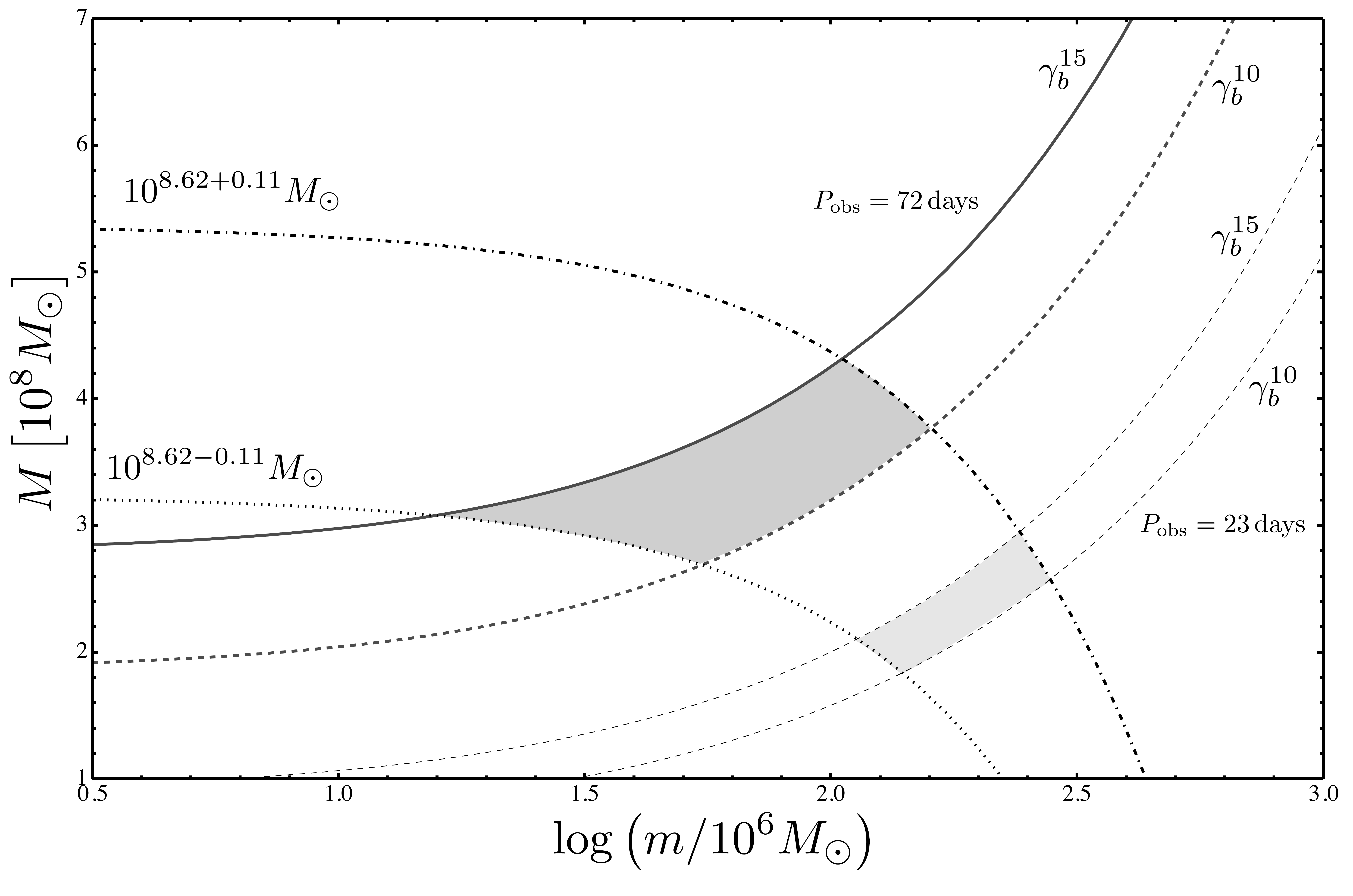}
  \caption{Required mass dependence for a BBH system in Mrk 501. The dotted and dot-dashed lines show the upper and lower limits of the binary mass estimation derived from the $M_{\bullet}-\sigma$ relation \citep{2005ApJ...631..762W}, respectively. The solid $(\gamma_{b} = 15)$ and dashed $(\gamma_{b} = 10)$ thick curves are given by the Doppler condition for inclination angles $i=1/\gamma_{b}$ and an observed period of $72$ days. The dashed thin curves represent the same relation for a period of $23$ days. The allowed mass-range has been indicated by filled areas. A TeV spectral index of $\alpha=1.2$ has been applied for the calculation.}
  \label{fig:massrangesw}
\end{figure}

\begin{table*}
  \begin{tabular}{|c c|c c|c c|c c|}
    \hline
    \multicolumn{8}{|c|}{Lomb-Scargle}       \\
    \hline
    RXTE          & $5\%-\,$     & SWIFT         & $5\%-\,$     & MAGIC         & $5\%-\,$      & WHIPPLE       & $5\%-\,$     \\
    Period [days] & level    & Period [days] & level & Period [days] & level &Period [days] & level \\
    \hline
    $72.2\pm 2.3$ & $0.074$      &               &              &               &                            &               &              \\
    $35.0\pm 2.2$ & $0.072$      & $34.6\pm 2.8$ &  $0.16$      &               &                          &               &              \\

    $22.9\pm 1.1$ & $0.036$      & $22.2\pm 1.0$ &  $0.057$     & $26.5\pm3.5$  & $1.4$                 & $21.6\pm 2.4$ & $0.82$       \\

    \hline
    \hline
    \multicolumn{8}{|c|}{SigSpec}        \\
    \hline
    RXTE          & $5\%-\,$     & SWIFT         & $5\%-\,$     & MAGIC         & $5\%-\,$      & WHIPPLE       & $5\%-\,$     \\
    Period [days] & level & Period [days] & level & Period [days] & level & Period [days] & level       \\
    \hline
    $72.6\pm 4.3$ & $0.14$       &               &              &               &                               &               &              \\
    $36.9\pm 2.3$ & $0.073$      & $32.7\pm 0.5$ &  $0.029$     & $36.6\pm4.1$  & $0.88$                   &               &              \\
    $21.8\pm 1.2$ & $0.039$      & $21.6\pm 0.3$ &  $0.019$     & $22.4\pm3.5$  & $1.5$                    & $21.1\pm 3.5$ & $1.1$        \\
    \hline
  \end{tabular}
  \caption{\label{tab:compfinal} The $5\% -$ significance levels of the corresponding hypothesis tests brought face to face with the identified 72 day period and its higher harmonics  }
\end{table*}

This study allows for a new limit on black hole masses in the
possible BBH system in Mrk 501. These limits are in
accordance with measurements with the total black hole mass derived
with other techniques.
For Mrk 501 the observed data corresponds to $f=8$ (where $f = S_\textrm{max}(\nu)/S_\textrm{min}(\nu)$ is the observed maximum to minimum ratio of the spectral flux density) and a TeV spectral index
of $\alpha \approx (1.2 - 1.7)$ \citep[cf.][]{1999A&A...349...11A}. Together with our proposed periodicity of $72$ days we can determine the masses of the constituting black holes of the binary system by simultaneously solving Eq.~(\ref{mass-ratio}) and (\ref{mass-ratio2}). The results are shown in Table~\ref{table:systemparameters}. In Fig.~\ref{fig:massrangesw} we show the limiting black hole masses for the large as well for the small black hole. These are compared for the new found
72-day period and the previously found 23-day period. 

The obtained value for the separation of the BBH obviously changes in time as a consequence of gravitational radiation. In Fig.~\ref{fig:massrangesw} we show the limiting black hole masses for the large as well for the small black hole. These are compared for the new found 72-day period and the previously found 23-day period. \\

%

\begin{table}[h]
  \begin{tabular}{|c|c|c|}
    \hline
    $i=1/\gamma_{b}$  & $1/10$          & $1/15$ \\
    \hline
    \hline
    $m \left[ 10^8 M_{\odot} \right]$ & 1.68       & 0.52       \\
    $M \left[ 10^8 M_{\odot} \right]$ &    2.49    &  3.65      \\
    $d \left[ 10^{16} \mathrm c\mathrm m \right]$ &  7.98      &   13.70     \\
    $P_{k} \left[\textrm{yrs} \right]$ &  19.11      &  42.96      \\
    \hline
  \end{tabular}
  \caption{ The maximum masses of the constituting black holes, the separation $d$ and the intrinsic orbital period for the inclination angles $i=1/\gamma_{b}$ with bulk Lorentz factor $\gamma_{b}$. A total mass of $10^{8.62} M_{\odot}$ \citep{2005ApJ...631..762W} and a TeV spectral index of $\alpha = 1.2$ have been applied for the calculations.}
    \label{table:systemparameters}
\end{table}


Our analysis shows two main results: Even for the quiescent state of the source Mrk501 we find that on the one hand the
well-established 23-day period \citep{osone} is still undefeated, while on the other
a 72-day period is upcoming. These results are not in
contradiction since the 23- and also the 36-day periods can, within the uncertainties, be regarded as
the second and first harmonic of the $72\pm 4.3$-day period, respectively. Fig.~\ref{rxtelomb} clearly
identifies the 72 days, together with the persistent appearance of the
first and second harmonics in the data sets (save for SWIFT, see
discussion above). It is not fully clear how robust the applied methods are by themselves, however the combination of different approaches showing concordant results provides a solid argument in favour of the SMBBH model.

This finding has possible consequences for our model of Mrk 501: It allows for a new limit on the black hole masses in the source's possible BBH system. A limit which is firstly in accordance with measurements of the total black hole mass as derived from other techniques and secondly allows for a larger ratio $r= M_{total}/m_{BH2}$  between the total mass of the binary and the mass of the companion BH. A scenario which is much more likely to exhibit the helical structure as proposed by \cite{rieger00}.

There is a reason why this has not yet been found: Short data sets are
not able to show the 72-day feature since their length has to be well
above these 72 days or aliasing will smear out this feature.

Regarding the physical model there is now, with a multi-wavelength
analysis, more reason to support the BBH model, but still
long time observations, especially in very high energies, are needed
to prove our results.

The results of this study have an impact on x-ray, gamma-ray and
gravitational wave astronomy: With evidence for multi-wavelength
periodic variability, gravitational wave telescopes may help to tackle
the problem of the origin of such variations and also aid in
restricting the parameters of these sources.

\begin{acknowledgements}
    The authors would like to thank the anonymous referee for his detailed comments, which greatly helped improving the paper.\\
    CR thanks Mr Piet Reegen for supplying her with the source of his SIGSPEC - code.
    CR acknowledges support from the Max-Weber-foundation, OE is grateful for funding from the Elite Network of Bavaria.
    TB thanks the Deutsches Zentrum f\"ur Luft- und Raumfahrt and the LISA-Germany collaboration for funding.
    This research has made use of data obtained through the High Energy Astrophysics Science Archive Research Center Online Service,
    provided by the NASA/Goddard Space Flight Center.
\end{acknowledgements}


\section*{}
\newpage
\bibliographystyle{aa}



\end{document}